%% file: ms.tex
\documentclass[12pt]{article}
\usepackage{epsf,aaspp4,flushrt}

\begin{document}

\title{HST IMAGING OF THE HOST GALAXIES OF 
HIGH REDSHIFT RADIO-LOUD QUASARS\altaffilmark{1}}

\bigskip
\bigskip
\author{Matthew D. Lehnert\altaffilmark{2,3}}
\affil{Sterrewacht Leiden, Postbus 9513, 2300 RA Leiden, The Netherlands}
\bigskip
\author{Wil J. M. van Breugel}
\affil{Institute of Geophysics \& Planetary Physics, Lawrence Livermore
National Laboratory, L-413 P.O. Box 808, Livermore, CA 94550}
\bigskip
\author{Timothy M. Heckman}
\affil{Department of Physics and Astronomy, 
Johns Hopkins University, Baltimore, MD 21218}
\medskip
\and
\medskip
\author{George K. Miley}
\affil{Sterrewacht Leiden, Postbus 9513, 2300 RA Leiden, The
Netherlands}

\medskip

\altaffiltext{1}{Based on observations with the NASA/ESA {\it
Hubble Space Telescope}, obtained at the Space Telescope Science
Institute, which is operated by the Association of Universities for
Research in Astronomy, Inc., under NASA contract NAS5-26555.}
\altaffiltext{2}{Participating Guest, Institute of Geophysics \&
Planetary Physics, Lawrence Livermore National Laboratory, L-413 P.O.
Box 808, Livermore, CA 94550}
\altaffiltext{3}{Current address: Max-Planck-Institut f\"ur extraterrestrische
Physik (MPE), Postfach 1603, D-85740 Garching, Germany, mlehnert@mpe.mpg.de}

\medskip

\vskip 0.1in
\centerline{Received ........................; accepted ........................}

\newpage

\begin{abstract}

We present rest-frame UV and Ly$\alpha$ images of spatially-resolved
structures (\lq hosts\rq) around five high-redshift radio-loud quasars
obtained with the WFPC2 camera on the Hubble Space Telescope.  The quasars were
imaged with the PC1 through the F555W (`V'-band) filter, which at the
redshifts of the quasars ($2.1 < z < 2.6$) have central wavelengths of
$\lambda_{rest}$ $\approx$ 1500\AA -- 1800\AA, and at rest-frame
Ly$\alpha$ using appropriately chosen narrow-band filters with the
WFC2.  The objects were selected from ground-based imaging surveys.
Those had shown that many radio loud quasars at high redshift have
prominent host galaxies which appeared to have properties similar to
those of high redshift radio galaxies.  Our HST observations allow a
more detailed investigation of quasar host morphologies and a
comparison with similar HST studies of radio galaxies by others.

Using several methods to measure and quantify the host properties we
find that all five quasars are extended and this ``fuzz'' contains
$\approx$ 5--40\% of the total continuum flux and 15--65\% of the
Ly$\alpha$ flux within a radius of about 1.5\arcsec.  The rest-frame UV
luminosities of the hosts are log $\lambda P_\lambda$ $\approx$ 11.9 to 12.5
$L_{\sun}$ (assuming no internal dust extinction), comparable to the
luminous radio galaxies at similar redshifts and a factor 10 higher
than both radio-quiet field galaxies at z $\sim$ 2 -- 3 and the most
UV-luminous low redshift starburst galaxies.  The Ly$\alpha$
luminosities of the hosts are log L$_{Ly\alpha}$ $\approx$ 44.3 -- 44.9
erg s$^{-1}$ which are also similar to the those of luminous high
redshift radio galaxies and considerably larger than the Ly$\alpha$
luminosities of high redshift field galaxies.  To generate the  Ly$\alpha$
luminosities of the hosts would require roughly a few percent of the total
observed ionizing luminosity of the quasar.

The UV continuum morphologies of the hosts appear complex and knotty at
the relatively high surface brightness levels of our exposures (about
24 V mags arcsec$^{-2}$).  In two quasars we find evidence for
foreground galaxies which confuse the host galaxy morphologies and
which are responsible for some of the perceived radio/optical
mis-alignments observed in ground-based imaging data.  We do find good
alignment between the extended Ly$\alpha$ and the radio sources, strong
evidence for jet-cloud interactions in two cases, again resembling
radio galaxies, and what is possibly the most luminous radio-UV
synchrotron jet in one of the hosts at z=2.110.  We discuss the
significance of jet-cloud collisions in radio-loud quasars and their
influence on radio morphologies in the frame work of simple
orientation-based quasar/radio galaxy unification schemes.

Our observations suggest that the host galaxies of radio-loud steep
spectrum quasars are similar to those of radio galaxies and strengthen
previous conclusions based on ground-based data that both types of
objects are probably members of the same parent population.

\end{abstract}


\newpage

\section{Introduction}

It is no exaggeration to say that between redshifts of $\approx$ 2-3 to
the present, highly luminous (M $<$ $-$26) quasars have become
virtually extinct.  The co-moving space density of luminous quasars has
fallen by about a factor of 1000 between these two epochs ({\it
e.g.,\ } Hartwick \& Schade 1990; Boyle 1993).  What processes led to
such a strong density evolution of highly luminous active galactic
nuclei (AGN)?  The answer to this question not only has important
ramifications for our understanding of the AGN phenomenon, but will
almost certainly give us important insight into the processes that
controlled the evolution of galaxies in general since the \lq\lq quasar
epoch\rq\rq .

It is an intriguing possibility that there is a strong link between the
fueling of luminous AGN and galaxy formation.  Indeed, the quasar epoch
occurs at the time \lq\lq Cold Dark Matter\rq\rq \ models identify as
when typical present-day galaxies where hierarchically assembled via
dissipative mergers.  Rees (1988), without regard to any specific
cosmogony, argues that the process of galaxy formation was still
continuing at the epoch z=2.  This is especially true of the
theoretical models where galaxy formation is a slow, diffuse,
low-luminosity process (e.g., Baron \& White 1987; Kauffmann, White, \&
Guiderdoni 1993; Baugh et al.~1998).  These general speculations
therefore give a special prominence to high redshift AGN by
hypothesizing that only when the proto-galactic material is energized
by the luminosity of an AGN will it have high enough surface brightness
to be readily detectable in emission-line surveys. In turn, the process
of galaxy formation may be substantially altered by the effect of the
AGN.

Observations of radio-loud quasars are also valuable for what we can
learn about the AGN phenomenon in general.  In particular, the
relationship between radio-loud quasars and radio galaxies is of
considerable interest, especially in light of efforts to \lq\lq
unify\rq\rq \ these two classes through differences in viewing angle,
environment, or evolutionary state ({\it e.g.,\ } Barthel 1989; Norman
\& Miley 1984; Neff \& Hutchings 1990).  In particular, the \lq\lq
viewing angle\rq\rq \ scheme of Barthel (1989) in which radio-loud
quasars and radio galaxies are drawn from the same parent population
but viewed preferentially at small or large angles, respectively, to
the radio axis, predicts that the luminosity and color of the quasar
host should be very similar, if not identical, to those of the radio
galaxies at similar radio powers and redshifts.  Moreover, radio
galaxies at high redshifts (z $>$ 0.6) exhibit the so-called \lq\lq
alignment effect\rq\rq \ (McCarthy et al.~1987; Chambers et al.~1987)
where the radio, and the rest-frame UV and optical axes are all roughly
co-linear.  Recently, through the use of broad-band HST imaging data,
it has become clear that quasars do indeed exhibit the ``alignment
effect'' (Lehnert et al.~1999a) and that the effect in quasars is
weaker than that seen in radio galaxies, as expected in simple
orientation-based unification schemes (Lehnert et al.~1999b in
preparation).

Radio-loud quasars at redshifts of 2-3 have impressively large and
luminous continuum and emission-line nebulae (``hosts''; Heckman et
al.~1991a,b; Lehnert et al.~1992).  These structures comprise about
3\% to over 40\% of the total flux from the quasar and have Ly$\alpha$
luminosities of $\approx$10$^{44-45}$ ergs s$^{-1}$, absolute visual
magnitudes of about $-$25, blue spectral energy distributions
consistent with those of  nearby late-type galaxies (Sc and Irr), and
sizes of the order of many tens of kiloparsecs.  Ground-based images of
high-z radio loud quasars contain only limited morphological
information (1 arc second seeing corresponds to about 11 kpc at these
redshifts) and their morphology can best be described as asymmetric.
Host galaxies of low-redshift quasars ({\it e.g.,\ } Boroson \& Oke
1984; Boroson, Persson, \& Oke 1985; Smith et al.~1986) have
emission-line luminosities much lower than this (about 10$^{42}$ ergs
s$^{-1}$ in [OII]), spectral energy distributions similar to late-type
galaxies, and absolute magnitudes of around $-$22.  However, Bahcall et
al.~(1994, 1995a,b) have recently concluded using HST WFPC2 images
that the host galaxies of some of the brightest, low-redshift QSOs have
wide range of host luminosities, with perhaps a large fraction of
quasars hosts having luminosities $< L_{\star}$.  However, subsequent
re-analysis of the data show that the host galaxies of low redshift
quasars are in fact large luminous galaxies (McLeod \& Rieke 1995;
Bahcall et al.~1997).  In spite of this, the properties of host
galaxies of quasars have clearly evolved very strongly from
high-redshifts (Kotilainen, Falomo, \& Spencer 1998; Lehnert et al.~ 1992).

As tantalizing as the data and information contained in these studies
of high redshift quasars are, there are some serious limitations to the
usefulness of our previously obtained ground-based data.  It is
essential to be able to separate the strong nuclear emission of the
quasar (continuum and Ly$\alpha$) from that of its surrounding host.
This can be most easily accomplished with the HST.  Accurate separation
of the quasar nucleus from its surrounding emission (within 10-20 kpc)
allows us to address a number of issues which are of importance to our
understanding of the evolution of luminous quasars, the association and
similarity between high-redshift radio galaxies and radio-loud quasars,
and the role of the large-scale radio emission in exciting and
enhancing the extended line emission.  To these ends, we used the HST
to study a small subset of high-redshift, radio-loud quasars from our
ground-based sample (Heckman et al.~1991a).

In this paper we represent an analysis of these observations, using
several different methods to test the robustness of our results, and
compare the quasar host properties with those of radio galaxies.
Throughout this paper we adopt H$_0 = 50$ km s$^{-1}$ Mpc$^{-1}$, q$_0
= 0.1$, and $\Lambda = 0$ for easy comparison to previously published
work on high redshift radio galaxies.

\section{Observations and Reduction}

The HST observations of our sample were made from September 1994
through May 1995 using the Wide Field Planetary Camera 2 (WFPC2).  Each
quasar was centered on the PC and was observed for a total of 2100
seconds (3 $\times$ 700 seconds) through the F555W filter (which
closely corresponds to a ground-based V filter).  In addition, we
obtained exposures of 5000 seconds (5 x 1000 seconds) using the Wide
Field Camera (WFC) and one of the Quad [OII] filters.  The narrow-band
filter was selected such that its central wavelength corresponded
closely to the wavelength of redshifted Ly$\alpha$ from each of the
quasars.  The observations are summarized in Table 1.  The pixel size
is 0.046\arcsec \ pixel$^{-1}$ in the PC and 0.1\arcsec \ pixel$^{-1}$
in the WFC.  The individual exposures were reduced using the standard
pipeline reduction.  The final images were produced by averaging the
individual exposures with sigma-clipping to remove cosmic rays.

The bandpass of the telescope + WFPC2 + F555W combination covers the
wavelength range of approximately 4500\AA \ and 6000\AA .  The strong
UV lines of CIV$\lambda$1550, HeII$\lambda$1640, and CIII]$\lambda$1909
are included in this bandpass and in principle it is possible that
these lines contribute to the detected extended emission.  However,
Heckman et al.~(1991$b$) and Lehnert \& Becker (1998) have shown that
these lines are generally weak in quasar hosts, with the possible
exception of PKS 0445+097. For this quasar Heckman et al.~(1991$b$)
found that the host has a HeII$\lambda$1640 equivalent width of several
\AA.  From this and the brightness of the host in our PC image we
estimate that HeII emission contributes much less than 1\% to the total
emission seen in the PC image.  Thus the contribution of line emission
to the total emission observed from the host in the F555W filter is
negligible, for PKS 0445+097 and all other quasars in our sample.

The broad-band data were flux-calibrated assuming the inverse sensitivity
for the F555W filter of 3.459$\times$ 10$^{-18}$ ergs s$^{-1}$
cm$^{-2}$ \AA$^{-1}$ dn$^{-1}$ and a zero point of 22.563 (Whitmore
1995).  This puts the resultant magnitudes on the ``Vega system''.  To convert to
the STMAG system, which assumes a flat spectral energy distribution and
has a zero point of 22.543, 0.020 magnitudes should be subtracted from
the magnitudes given here.

\section{Image Reduction}

\subsection{Construction and Systematics of the PSF}

We have attempted to quantify the shape and constancy of the HST PSF.
First, we have collected images of the standard stars used to calibrate
the F555W filter and observations of bright stars in the outer regions
of $\omega$ Cen that were made within a period of days of our
observations.  We found about 20 stars that were suitably close in time
and within about 100 pixels of the central pixel of the PC (all of
quasar images where centered near the middle of the PC CCD).  We then
constructed an empirical point-spread function using these data by
adding the individual exposures after they had been aligned to a common
center.  This empirical PSF was then compared with the model PSF
constructed using the PSF modeling program, ``Tiny Tim''.  However, the
close agreement is limited to azimuthal averages -- Tiny Tim does not
reproduce the detailed 2-dimensional structure of the PSF (there is an
asymmetry in the intensity of the diffraction spikes, especially in the
$+$U3 direction, which Tiny Tim does not reproduce well).  Next, we
also measured the encircled energy diagrams (EEDs), i.e., the fraction
of flux from a point source interior to a radius r, as a function of r.
We then inter-compared all the EEDs taken through a given filter to
determine the reproducibility of the EED and compared the individual
stellar EEDs with that of the model PSF produced by Tiny Tim.  We found
very good agreement between the shape of the EED from the sum of the
observations of standard stars and that of the Tiny Tim PSF.  We then
compared individual star EEDs with that of the Tiny Tim PSF.  This
inter-comparison of approximately 20 stars showed that we can detect
host that contributes more than about 5\% as much light as the quasar
itself (within a radius of about 1.4\arcsec).  This limit is consistent
with the known temporal variations in the HST PSF due to effects like
the gentle change in focus over times scales of months and shorter time
scale variations due to the so-called ``breathing'' of the telescope
(see Burrows et al.~1995).  We have restricted ourselves to radii less
than about 1.5\arcsec , to avoid the effect of the poorly understood
large angle scattering which becomes important beyond a radius of about
2\arcsec.

\subsection{Host Measurements: PSF Subtraction}

Measuring the properties of the circum-nuclear emission is important if
we are to gain true insight into the properties of quasar host
galaxies.  One arc second at the redshift of quasars corresponds to
roughly 11 kpc (in the cosmology: H$_0$=50 km s$^{-1}$ Mpc$^{-1}$ and
q$_0$=0.1 which will be used throughout this paper) -- which is similar
to the sizes of present-day galaxies.  Thus being able to accurately
subtract the nuclear emission of the quasar from the more extended
emission is critical if we are to obtain information about the host of
high redshift quasar on scales a fraction of the present-day galaxy
size.  We attempted this subtraction using two different PSFs.  In the
first, we modeled the PSF using the HST PSF modeling program, Tiny
Tim.  The second method was to construct a PSF by averaging several
images (40s integrations) of a bright star in the outer regions of
$\omega$ Centauri using the F555W and the PC.  These stellar images
were all within about 100 pixels of the center of the PC chip and thus
near the location of the quasar image on the PC chip.  The images were
aligned to the nearest pixel before averaging and the individual images
covered the central 200 pixels of the planetary camera.  The counts in
the central pixel was about 2500 DN, comparable to that in the images
of the quasars.  We restricted ourselves to these observations and not
the entire 20 that were used in the EED analysis.  The subsample we
used were chosen to sample the range of PC positions covered by the
F555W quasar exposures as part of this program.

These two PSFs were iteratively subtracted until emission due to the
diffraction of the secondary support was zero.  This procedure allowed
us to estimate the uncertainty in our fraction of extended emission by
observing the points were the diffraction spikes became negative due to
over subtraction of the PSF or were obvious in the PSF subtracted image
due to under subtraction.  We estimate that the range of possible acceptable
amounts of PSF subtraction lead to a factor of 30\% or $\pm$0.3
magnitudes in the quoted flux from the host.  We note that these
diffraction spikes were seen only out to about 1\arcsec \ in the PC frames and
there was very little underlying nuclear emission from the quasar over
most of this area.  Using the model from Tiny Tim or the images of the
star in $\omega$ Cen gave very similar results ({\it i.e.,\ } very
similar fraction of the emission that was extended and similar
morphology of the underlying emission).  Therefore throughout the
remainder of the paper, we will quote only the results of PSF
subtraction obtained by subtracting the image of the star through the
F555W and the PC.

The narrow-band images of the quasars generally only reveal modest
extended emission.  Using the efficiencies of the telescope, detector
and filter combination made available by STScI and the integration
time, we calculate that the scaling factor necessary to remove the
continuum contribution to the narrow-band filter using the F555W filter
to be about 50 -- 100.  Therefore continuum subtraction has only
negligible effect on the properties of the extended emission.
Unfortunately, we could not empirically measure this scaling factor.
This is because the Quad [OII] filters only cover one of the WFCs at a
time and in each of the images of the quasars, the stars available for
such an estimate were either too faint to be useful, or were too bright
and were saturated in the F555W image.  To continuum subtract the
narrow-band data, we block-averaged the PC continuum images 2 $\times$
2 to make the scale of the PC continuum images match that of the WFC
narrow--band images.  The two images were then aligned and the F555W
image then scaled by the factor calculated above.  The final image is
referred to as the Ly$\alpha$ image.  The fluxes measured in the
Ly$\alpha$ images are in reasonable agreement (always within a factor
of 2) with those obtained from the ground by Heckman et al.~(1991a).

The number of stars available for the construction of empirical PSF for
the narrow-band frames was very limited.  However, each individual star
used for the construction had many times the counts in the images of
the quasars and therefore, even with just a few stars, the signal to
noise in the final PSF was sufficient to make a reasonable subtraction
of the PSF.  We note that in none of the cases did a strong point
source appear in the images and thus the method of PSF subtraction was
a little bit different from the subtraction of the continuum images.
None of the narrow-band images of the quasars had obviously visible
diffraction spikes and thus most of the extended emission is little
effected by the structure of the PSF.  We therefore subtracted until
the emission from the central peak roughly blended in with the more
extended emission ({\it i.e.,\ } the subtraction did not cause a hole
in the position of the nucleus).  In Table 2, we quote the results of
the PSF subtraction.

\subsection{Host Measurements: Other Methods}

We attempted several other techniques for estimating the amount of
extended emission from each quasar.  One relied on scaling the PSF such
that its flux in the central 2 pixels matched that of the quasar
image.  We then took the ratio of the total energy encircled in
apertures of increasing radius.  This method provides an estimate of
the minimum amount of extended flux in each quasar.  The other
technique we used to make this estimate was to cross-correlate a series
of galaxy models plus PSF (to represent the underlying galaxy and
quasar) with the image of the quasar.  This technique was developed in
a series of papers (Phillips \& Davies 1991; Boyce, Phillips, \& Davies
1993).  Complete descriptions of the technique and its robustness can
be found in the original papers cited above.  We made this comparison
with a series of model elliptical galaxies of varying half-light radii
and ellipticities.  Since we and the referee found that given the
complexity of the host morphology revealed in the point-spread-function
subtracted images, these other methods were perhaps not as
convincing as the PSF subtraction.  Thus, we will not go into detail of the
results of these other methods.  It is sufficient to say that they gave
estimates very similar to those obtained through PSF subtraction.

\section{Results}

Contour plots of the HST images of all 5 quasars are shown in Figure
1.  To compare the optical and radio structures we also overlaid high
resolution radio maps from Lonsdale, Barthel, \& Miley (1993) on the
HST images, as shown in Figure 2.  For these overlays, the coordinates
from the HST images are insufficiently accurate to directly overlay the
images using the coordinates given by the standard pipeline reduction.
There is an uncertainty of 0.5\arcsec \ to 1\arcsec \ between the
absolute HST positions as given by the pipeline reduction and the radio
coordinate system.  Therefore, we made some assumptions about how to
position the radio images relative to the HST images.  Using published
spectral information available for these quasars, we identified the
flattest spectrum component in each radio image, which is presumably
associated with the radio AGN, and aligned that component with the most
intense pixel ({\it i.e.,\ } the optical quasar nucleus) in the HST
image.  Second, we checked that our core determined using the method
just outlined agreed with the astrometry from Barthel (1984) between
high resolution VLA radio maps and the position of the quasar and in
every case we found good agreement.

\subsection{Notes Individual Objects}

\subsubsection{PKS 0445+097}

The continuum image of PKS 0445+097 is rather asymmetric and one can
discern two components:  an asymmetric circum-nuclear host within
$\sim$ 1\arcsec \ of the nucleus, elongated along P.A. $\approx$
240$^\circ$, and a detached ``blob'' at $\sim$ 2\arcsec \ from the
nucleus at P.A. $\approx$ 120$^\circ$  The host has a total magnitude
of about 22.1$\pm$0.3 magnitudes through the F555W filter.  This
``blob'', located about 1.5\arcsec \ ($\sim$17 kpc) to the east-southeast of the
quasar nucleus, has a total magnitude of 23.1$\pm$0.2 and is composed
of several bright clumps immersed in a more diffuse structure.  At
surface brightness levels of $\approx$23 m$_{F555W}$ arcsec$^{-2}$, its
total extent is approximately 2.5 $\times$ 1.5 arc seconds (along PAs
of 35$^\circ \pm$10$^\circ$ and 125$^\circ\pm$20$^\circ$).  There is
hint of a faint tail of emission leading from the region about the
quasar nucleus out toward this blob of emission to the east-southeast.
Another interesting feature in the circum-nuclear continuum emission is
a $\sim$1\arcsec \ long ``arc'' of emission that curves out eastward of the
nucleus and bends to the southeast.  The total extent of the
circum-nuclear emission, down to surface brightness levels of 23
m$_{F555W}$ arcsec$^{-2}$, is about 2.5 $\times$ 0.8 arc seconds
(roughly along PA = 105$^\circ$ and PA = 15$^\circ$; 30 kpc $\times$ 9
kpc for z=2.110).  The total extended emission (both the circum-nuclear
host and blob to the southeast) comprises about 25\% of the total
emission seen from PKS 0445+097 in the F555W filter.  The
circum-nuclear host contributes about 17\% of the total and the blob to
the southeast of the nucleus about 7\%.

The F555W continuum image is qualitatively and quantitatively
consistent with the broad-band continuum images from Lehnert et
al.~(1992).  These ground-based images show that the extended emission
is blue (consistent with that of a nearby Irregular or Sc galaxy) and
extended on spatial scales of about 8 arc seconds ($\sim$100 kpc) to
the southeast (Lehnert et al.~1992).  The HST image does not show as
large of an extent, only about 2 -- 3 arc seconds, but the orientation
of the host is similar and the relative fraction of extended emission
is the similar (15 -- 40\% in the ground-based data).  The HST image is
more sensitive to the high surface brightness and more compact
structures while the ground-based images with their inferior spatial
resolution and larger projected pixel sizes, are more sensitive to the
low surface brightness and more diffuse emission.  This being the case,
inspection of the HST F555W and our previously published ground-based
images (Lehnert et al.~1992) suggests that both high surface brightness
areas -- the circum-nuclear host and the blob of emission -- are
embedded in a diffuse area of emission on scales of tens of kpc
preferentially oriented along the axis southeast to northwest but also
having some low surface brightness to the north, northeast, and
southwest of the nucleus.

The narrow-band image of PKS 0445+097 is peculiar.  The flux of the
brightest source in the image is lower by a factor of $\sim$100 than
what we would have expected compared to the flux measured in a
ground-based image (Heckman et al.~1991a).  The filter we used for the
narrow-band imaging of PKS 0445+097 is unusual in one respect compared
to the other filters.  In the filter holder, the FQUVN-A filter is
located within the beam of the planetary camera.  To move the filter
onto one of the wide-field camera arrays, required the filter holder to
be rotated 33$^\circ$.  We attribute the discrepancy in the measured
flux to an unknown error in the positioning of the filter or of the
target and for the remainder of the paper, we will not consider the
narrow-band image of PKS 0445+097 further.

\subsubsection{MRC 0549-213}

The continuum image of MRC 0549-213 reveals a complex structure
surrounding the quasar nucleus.  The emission is symmetric about the
nucleus, with the principal axis of the emission changing from about
90$^\circ$ within a few tenths of an arc second of the position of the
nucleus, to a position angle of about 150$^\circ$ at a distance of
1\arcsec .  The total magnitude of the quasar (nucleus plus host) is
about 19.7 and the magnitude of the host is about 21.3$\pm$0.3
magnitudes in the F555W filter.  The fraction of the total brightness
contributed by the host is about 23\%.  In addition, we see a complex
structure of continuum emission about 3.2 arc seconds to the west of
the nucleus.  This emission region has a curved arc shape and is
approximately 1\arcsec \ in size, down to surface brightness levels of
$\approx$23 m$_{F555W}$ arcsec$^{-2}$.  It has a total magnitude of
about 24.0.

The F555W continuum image is qualitatively and quantitatively
consistent with the U band continuum image in Heckman et al.~(1991).
The ground-based U image shows that host is very extended, with the
near-nuclear emission (within a few arc seconds of the nucleus) being
preferentially oriented along PA $\approx$ 150$^\circ$.  Moreover,
there is a ``tail'' of emission that extends about 5 arc seconds to the
west.  This morphology agrees quite well with that seen in our F555W
continuum image.  Comparing the ground-based and HST image in detail
suggests that the emission from the near-nuclear environment of the
quasar and the blob to the west must actually be physically connected
(deep R-band images of MRC 0549-213 obtained as part of another
ground-based program shows a very similar morphology).  Our HST image
is not sufficiently deep to detect this connection.  Also, the
ground-based data suggested that about 20\% of the U band flux is
extended, consistent with the 23\% estimated from our HST data.

The narrow-band image of MRC 0549-213 shows extended structure.  Within
a few tenths of an arc second of the nucleus, the emission is extended along
PA$\approx$ 45$^\circ$.  On a scale of a few arc seconds there are two
regions of significant emission.  One is along PA $\approx$170$^\circ$
and extends out about an arc second from the nucleus.  About 1.6 arc
second to the east of the quasar nucleus, there is a faint blob of
emission that seems to be connected to the quasar nucleus proper.  This
faint blob of Ly$\alpha$ emission is approximately 1 arc second long in
the north-south direction and about 0.5 arc seconds wide in the
east-west direction down to our detection limit. The flux from this
distinct region of emission is 9.7$\times$ 10$^{-16}$ ergs s$^{-1}$
cm$^{-2}$.  In addition, we see several areas of low surface brightness
emission near this object.  Two regions are particularly noteworthy.
One area corresponds to the blob 3.2 arc seconds to the west.  There is
a $\approx$ 4 $\sigma$ region of Ly$\alpha$ emission over the region of
this blob.  Also, there is some Ly$\alpha$ emission roughly
corresponding to a area of continuum emission about 3 arc seconds away
from the nucleus along PA $\approx$ 325$^\circ$.  Unfortunately, there
does not exist a ground-based Ly$\alpha$ image of this quasar with
which to compare.

\subsubsection{PKS 1318+113}

The continuum image of PKS 1318+113 shows two concentrations of
emission, one immediately surrounding the quasar nucleus and the other
about 2 arc seconds to the east of the quasar nucleus.  Immediately
surrounding the quasar nucleus (within about 1\arcsec), the emission is
asymmetric, with the brighter isophotes oriented along PA $\approx$
135$^\circ$ and the fainter isophotes are most extended along PA
$\approx$ 180$^\circ$ to 200$^\circ$.  The total magnitude of the
quasar is about 19.0 in the F555W filter and the host has a magnitude
of about 20.1.  This implies that the host makes up about 38\% of the
total emission from the quasar (nucleus + host).  We note that
perhaps this is somewhat over-estimated in light of the fact that the
cross-correlation technique implies that only about 19\% of the quasar
light is extended.  The galaxy to the east of the nucleus has a
magnitude of about 21.9 (measured in a 2\arcsec \ $\times$ 2\arcsec \ box, which is as large as can
be used due to the proximity of this galaxy to the line-of-sight of the
quasar).  Down to surface brightness levels of
$\approx$24 m$_{F555W}$ arcsec$^{-2}$, the extent of the circum nuclear
host is about 1 to 1.5 arc seconds.

The HST F555W image is similar to the B-band image presented from
Heckman et al.~(1991a).   The ground-based image shows bright extended
emission to the east of the nucleus, with fainter emission to the south
and west.  The total extent of the ground-based B image is about 6 -
10\arcsec \ from the nucleus.  The HST image does not reveal emission quite as
extended as this, only about 2 -- 3 arc seconds, but the gross
characteristics of the host is similar.  The relative fraction of
extended emission between the ground-based and HST data are not very
similar (16\% in the ground-based B data versus about 38\% in the HST
F555W data; although we note that the cross-correlation analysis gives
19\%; see Table 2).  This suggests that relatively speaking, the light
from the host is concentrated within an 1\arcsec \ of the nucleus (scales not
available from the ground).  Although again, we note that the
cross-correlation analysis gives a result much more consistent with our
previous ground-based results.

The narrow-band image of PKS 1318+113 shows extended emission (Figure
1).  Most of the extended Ly$\alpha$ emission is to the north and east
of the nucleus, primarily along PA$\approx$ 45$^\circ$ and is extended
on 1\arcsec \ to 2\arcsec \ from the nucleus (down to surface brightnesses of 6.3
$\times$ 10$^{-16}$ ergs s$^{-1}$ cm$^{-2}$ arcsec$^{-2}$.  There are
several faint regions of emissions within a few arc seconds of the
quasar.  These regions have fluxes of roughly 1.7 to 7.1 $\times$
10$^{-16}$ ergs s$^{-1}$ cm$^{-2}$.  Moreover, we find reasonable
agreement with the morphology of the ground-based Ly$\alpha$ image
presented in Heckman et al.~(1991a).  In the ground based Ly$\alpha$
images the host was extended along PA$\approx$45$^\circ$ with the most
extended emission being on the southwest side of the nucleus.  Detailed
comparison between the ground-based and HST Ly$\alpha$ images suggest
that the highest surface brightness emission is on the northeast side
of the nucleus with several bright clumps of the southwest side that is
then embedded in a halo of diffuse Ly$\alpha$ emission.

One of the most remarkable results of this small HST survey of the host
galaxies of high redshift quasars is the interesting spatial
relationship between extended Ly$\alpha$ and radio jet emission.  In
Figure 2, we show an overlay of the  Ly$\alpha$ HST image and a VLA
A-array map from Lonsdale et al.~(1993).  We see that the jet passes
between two Ly$\alpha$ emitting blobs southwest of the quasar nucleus.
This interaction appears at the point where the jet appears to bend.
The two blobs have total fluxes of 1.67$\times$10$^{-16}$ ergs s$^{-1}$
cm$^{-2}$ and 2.01$\times$10$^{-16}$ ergs s cm$^{-2}$ for the
eastern-most and western-most emission regions.  Measuring the sizes of
these blobs, we find that the eastern most blob is approximately
circular with a diameter of 0.3\arcsec.  The western most blob of the two is
approximately 1\arcsec \ $\times$0.35\arcsec \ (long versus short axis oriented
PA$\approx$150$^\circ$).  We also note that there seems to be another
region of Ly$\alpha$ emission along the ``counter-jet'' side of the
quasar between the nucleus and the north-eastern radio lobe.

\subsubsection{1658+575 (4C 57.29)}

The continuum image of 1658+575 (4C 57.29) shows a relatively compact
(about 1\arcsec \ across) region of extended emission.  The total magnitude of
the quasar (nucleus + host) is 18.3 and the magnitude of the host is
20.0.  We find that about 21\% of the total emission from the quasar is
extended.  We see a linear feature along PA$\approx$150$^\circ$ in the
extended emission.  This feature is very likely to be a residual
emission that was not accounted for during PSF subtraction.  This is
not surprising since 1658+575 is the brightest quasar imaged during
this program and hence had the most extended and intense diffraction
spikes compared to the other quasar images. Ignoring the linear feature
we see that the brightest emission is north and to the southwest of the
nucleus.  The diameter of the host is only about 1\arcsec \ down to surface
brightnesses of 22 m$_{F555W}$ arcsec$^{-2}$.

The ground-based B band image of 1658+575 presented in Heckman et
al.~(1991a) shows a structure roughly similar to that observed using
the HST.  The quasar is extended on scales of about 10\arcsec \ in the
ground-based image and has a high surface brightness region to the
east-southeast of the nucleus with lower surface brightness emission
also to the north-northeast and south of the nucleus.

The narrow-band image of 1658+575 shows an extended plume of emission
to the northeast with some very significant emission also extended to
the northwest of the nucleus.  There is also a lower surface brightness
extension to the south of the nucleus.  The most extended emission
reaches a radius of about 2\arcsec \ from the nucleus.  Comparison with the
ground-based Ly$\alpha$ image of Heckman et al.~(1991a) again reveals a
very strong similarity between the two images.  In the ground-based
data Ly$\alpha$ is extended on scales of up to 6\arcsec \ from the nucleus.
The most significant of this extended emission is to the northwest
through the south side of the nucleus.

\subsubsection{PKS 2338+042}

The host galaxy of the quasar PKS 2338+042 comprises nearly 40\% of
the total continuum emission from the quasar.  The host is asymmetric
with an ``arm'' of emission that emanates from the nucleus to the
south and then bends around to the east.  In addition, there
is a ``plume'' of emission to the northeast of the nucleus.  The
total extent of the continuum nebula is about 1.5\arcsec \ down to
a surface brightness of 24.0 m$_{F555W}$ arcsec$^{-2}$.  The circular
contour seen 0.4\arcsec \ to the east of the nucleus in the contour plot
is a local minimum in the emission.

The 15 GHz radio map of Lonsdale et al.~(1993) shows a ``bent'' core,
jet, double lobe source oriented preferentially in an east-west
direction.  The ``jet'' emanates from the nucleus along
PA$\approx$90$^\circ$, with the hotspot of the eastern lobe is at
PA$\approx$120$^\circ$.  The western hotspot is at
PA$\approx$270$^\circ$.  Overlaying this map onto the PSF subtracted
HST image, we see a close correspondence between features in the HST
image and that of the radio image.  The maximum intensity seen in the
contour plot of the radio emission just to the east of the nucleus
along the ``jet'' corresponds to the local minimum we see in the
contour plot of the F555W image.  This local minimum gives the
impression that we are looking down the end of a hollow tube of
emission which contains the radio emission.  Farther to the east we see
that the surface brightness of the continuum emission increases at
roughly the same point where the jet seems to ``bend'' towards the
hotspot.  Moreover, we notice that the rest-frame UV isophotes of the
F555W image seem to bend outwards to the west of the nucleus
approximately along the same position angle as that to the western
radio lobe ({\it i.e.,\ } PA$\approx$270$^\circ$).

The F555W continuum image is roughly consistent with the ground-based 4m
image taken through the B-filter by Heckman et al.~(1991).  This
ground-based image shows the most significant emission is to the
southeast of the nucleus and is extended about 6\arcsec \ across down to a
surface brightness of 27.3 m$_B$ arcsec$^{-2}$.  The HST image does not
reveal emission quite as extended as this, only a few arc
seconds, but the orientation of the host is roughly similar.  The HST
image however suggests that a much greater fraction of the emission is
extended, namely $\sim$40\% versus $\sim$16\% from the ground-based B
image.  The HST image is obviously more sensitive to the high surface
brightness and more compact structures while the ground-based images
with their inferior spatial resolution and larger projected pixel
sizes, are more sensitive to the low surface brightness and more
diffuse emission.  This would suggest that perhaps Heckman et al.~(1991)
over-subtracted the ground-based image of PKS 2338+042 and that
the UV continuum host of PKS 2338+042 is very compact compared to the
rest of this small sample.

The Ly$\alpha$ image of PKS 2338+042 is also very extended, revealing a
host galaxy approximately 2\arcsec \ across down to a surface brightness of
5.4 $\times$ 10$^{-16}$ ergs s$^{-1}$ cm$^{-2}$ arcsec$^{-2}$.  The
morphology of the host galaxy in Ly$\alpha$ is similar to that of the
continuum emission.  The basic orientation of the Ly$\alpha$ host is
east-west.  The most significant piece of the morphology of this image
is an ``arm-like'' structure that extends from the nucleus out to about
1 arc second to the east where it terminates in a relatively high.
surface brightness region of emission.  To the north of this ``arm''
there is a region of relatively low surface brightness compared to its
immediate surroundings (a ``hole'' in the extended emission).  There
are fainter ``plumes'' of emission to the northeast and the
south-southeast.  At the lowest surface brightness visible in the
Ly$\alpha$ image, there is also a faint extension to the northwest.  At
the lowest surface brightness levels, the orientation of the Ly$\alpha$
is preferentially in the southeast-northwest direction (PA$\approx$
140$^\circ$) as opposed to the general east-west orientation at higher
surface brightnesses.

The morphology of the HST Ly$\alpha$ image is similar to that seen in
the ground-based Ly$\alpha$ image of Heckman et al.~(1991).  The
ground-based Ly$\alpha$ image shows a general southeast-northwest
orientation with Ly$\alpha$ emission being seen over about 9\arcsec.  The
ground-based Ly$\alpha$ image is of course a much deeper image than our HST
Ly$\alpha$ image, reaching down to a surface brightness of 1.5 $\times$
10$^{-17}$ ergs s$^{-1}$ cm$^{-2}$ arcsec$^{-2}$.

Overlaying the 15 GHz radio image of Lonsdale et al.~(1993) on the HST
Ly$\alpha$ image we again see a good correspondence between the highest
surface brightness Ly$\alpha$ emission and the radio ``jet''.  The
Ly$\alpha$ shows a high surface brightness extension about 0.8\arcsec \
to the east of the nucleus.  Over this same region the jet of radio
emission is observed.  It is interesting that over the region of the
most intense extended Ly$\alpha$ emission is also the region where we
see the ``jet'' of radio emission and where the radio emission
undergoes its most severe bending (in projection).  Moreover, we again
see that to the west of the nucleus, the isophotes ``bend'' outwards
from the nucleus over a region about 0.5\arcsec \ out from the nucleus.  This
is along the same position angle from the nucleus that we see the most
distant radio hotspot

To elucidate the relationship between the radio and UV continuum and
Ly$\alpha$ line emission, we have made a single cut through the 15 GHz
radio image from  Lonsdale et al.~(1993), and both the PSF subtracted
F555W and narrow-band Ly$\alpha$ image (which has not been PSF
subtracted).  These cuts were made from the highest surface brightness
peak in the nucleus of the radio image, in the PSF subtracted F555W,
and in the Ly$\alpha$ image and then including all the pixels to the
east and west of the nucleus out to a radius of roughly an arc second
in both directions.  We have normalized and overlayed these cuts in
Figure 3.  The direction of the cut was selected such that it passed
directly along the radio jet that points directly to the east of the
nucleus so that we may directly compare the one-dimensional spatially
extended radio, UV continuum, and Ly$\alpha$ radial distributions.

As can be seen in this Figure, the radio and UV continuum emission are
strongly correlated, while the radio and Ly$\alpha$ emission are not.
Since this projected cut lies along the direction of the minimum in the
spatial distribution of the UV continuum (see Figure 1), the top panel
of Figure 3 shows that the jet must pass through the local minimum
(best described as a ``hole'') in the distribution of the UV continuum
emission.  On the other hand, the anti-coincidence of the radio and
Ly$\alpha$ distribution is such that at the position where the jet is
bending away to the southeast, which is why the radio surface
brightness in the bottom panel of Figure 3 is decreasing, the
Ly$\alpha$ surface brightness is increasing and reaching a local
maximum.  The spatial relationship between the Ly$\alpha$ and radio
emission is very suggestive of a ``jet-cloud'' interaction in that
the area of high surface bright Ly$\alpha$ emission is responsible for
bending the jet.

While the structure of the Ly$\alpha$ image is rather complex, to aid
us in interpreting the data in relationship to a possible ``jet-cloud''
interaction, we wish to estimate the flux from the region of relatively
high surface brightness in the Ly$\alpha$ image at the point where the
jet bends to the southeast.  Isolating the pixels over this region
(approximately 0.3\arcsec \ $\times$ 0.3\arcsec \ region about
0.8\arcsec \ from the nucleus), we find a total Ly$\alpha$ flux of 2.1
$\times$10$^{-16}$ ergs s$^{-1}$.

\section{Discussion}

In this section we discuss the results and their implications for our
understanding of the circum-nuclear environments of high redshift
quasars.  Our sample is too small for a detailed statistical analysis.
Thus, we will focus our attention on a few commonalities shared by the
quasar hosts and compare these properties with those of high redshift
radio galaxies and field galaxies, and low redshift starburst galaxies.

\subsection{The Radio-Aligned UV Continuum and Confusion by Intervening 
Absorber Galaxies}

Heckman et al.~(1991a) and Lehnert et al.~(1992), from ground-based
images of quasars, found weak evidence for alignment between the rest-frame
optical/UV and the radio emission.  In a HST/WFPC2 snapshot study of 43
quasars from the 3CR catalog, Lehnert et al.~(1999b) argue that
quasars hosts indeed exhibit the ``alignment effect'' in the continuum
plus line emission (all of the broad-band HST images in that study have
some contribution due to emission lines) but that the effect is
slightly weaker than in radio galaxies at similar redshifts.  As
discussed below, the HST data show that intervening galaxies along the
line of sight to the quasar may confuse the morphologies of the hosts:
2 of the 5 quasars from our sample appear to have nearby galaxies seen
in projection (PKS 0445+097 and PKS 1318+113).  Interestingly, these
two quasars were precisely those which showed the greatest
mis-alignment between the rest-frame UV continuum host and radio
emission in the study of Heckman et al.~(1991a).

The evidence that PKS 0445+097 has a nearby intervening system SE of
the nucleus is based on Keck 10m spectroscopy (Lehnert \& Becker 1998),
as well as morphological and luminosity considerations.  The Keck
spectrum shows that the SE blob is at a redshift of 0.8384$\pm$0.0002,
which is similar to that of the MgII absorption seen against the
nuclear continuum of PKS 0445+097 (Barthel, Tytler, \& Thompson 1990).
Therefore, it is not surprising that we found no Ly$\alpha$ emission
from this ``blob'' at the redshift of the quasar.  In addition, Lehnert
\& Becker also found that this galaxy is likely to contain a Seyfert 2
nucleus.  If we adopt z=0.84 for the redshift of this galaxy, the
central wavelength of the F555W filter corresponds to about 2930\AA
\ in the rest-frame of the galaxy.  This wavelength is close to the
wavelengths of the U and B filters and thus extrapolations from flux
density measured in the F555W to estimate the flux densities of the U
and B filters in the rest-frame of the quasar host are small.  Using
the spectral energy distribution from Armus et al.~(1997) for this
galaxy (approximately that of the a late-type spiral) to extrapolate
the measured flux density in the F555W filter to the flux density at
the wavelengths of the U and B filters in the rest-frame of the
intervening galaxy and correcting for Galactic extinction, we find that
the U and B absolute magnitude of the blob to the SE of the nucleus is
M$_U$ = $-$21.8 and M$_B$ = $-$21.3.  Thus this intervening galaxy is
approximately a factor of 2 more luminous than a fiducial Schecter
$L^{\ast}$ \ galaxy.

Moreover, this intervening galaxy appears to have a very distorted
morphology.  The HST F555W image shows a galaxy with three knots of
emission elongated along PA$\approx$45$^\circ$ with the brightest
region not roughly in the center of the emission but towards the
northeastern end of the galaxy.  The galaxy appears to be nearly
edge-on.  Comparing the morphology of this galaxy with other
intervening absorbers observed with the HST ({\it e.g.,\ } Dickinson \&
Steidel 1996; Steidel et al.~1997), we find that the galaxy
along the line of sight to PKS 0445+097 is peculiar.  Most MgII
absorbing galaxy have properties consistent with the general field
population of galaxies and hence ``normal'' morphologies and
distributions of luminosity similar to field galaxies (Steidel et al.~
1997; Bergeron \& Boisse~1991; Steidel et al.~1994).  However, many of
the peculiar morphologies appear to be associated with galaxies that
are viewed nearly edge-on (Dickinson \& Steidel 1996).  Thus we
conclude that even though the morphology appears to be peculiar
compared to most MgII absorbing galaxies, its apparently edge-on
orientation implies that extinction in the disk may account for its
seemingly peculiar morphology.  The two color analysis of Armus et
al.~(1997) suggests that the color of this galaxy is consistent with Sc
spiral galaxy at z=0.84 with about 0.5 magnitudes of addition
extinction compared to low redshift Sc spiral galaxies.  This
additional reddening is consistent with our claim here that extinction
may account for this galaxy's seemingly peculiar morphology.  However,
it could also be that since this galaxy appears to be harboring a
Seyfert nucleus (Lehnert \& Becker 1998), it might also be that the
complex morphology is associated with a merger event that initiated the
Seyfert activity.

In the case of PKS 1318+113 no direct spectroscopic evidence exists
that its companion to the east is also a foreground object.  If this
object is at the redshift of the quasar, it would have an implausibly
high luminosity ($>$ 25 $L^{\ast}$ and more luminous than the quasar
host).  There are several other more plausible possibilities for the
redshift of this object.  The two most plausible redshifts for this
object, 0.8388 and 1.0541, which are associated with Mg II absorbers
along the line of sight to PKS 1318+113 (Barthel et al.~1990;
Jankarkarinen, Hewitt, \& Burbidge 1991).  Since this galaxy is bright
(about 21.9 in F555W), it is more likely that this galaxy is associated
with the MgII absorber at z=0.8388, which is very similar to the case
of PKS 0445+097.  The absolute magnitude of this galaxy under the same
assumptions made previously for the absorber along the line of sight to
0445+097 implies M$_U$ $\approx$$-$23 and M$_B$ $\approx$$-$22.5.
These magnitudes are many $L^{\ast}$ and thus we consider associating
this galaxy with the MgII absorber at z=0.8388 very implausible and
that it is associated with the absorber at z=1.0541 even less likely
(see e.g., Bergeron \& Boisse~1991; Steidel et al.~1994).  However,
such a speculation will have to await spectroscopic observations to
determine the redshift of this nearby (in projection) galaxy.

\subsection{The Nature of the UV Continuum}

We will center our discussion of the UV continuum in the hosts 
of quasars on two aspects: the origin of radio-aligned UV continuum
and the stellar population of the underlying galaxy.

\subsubsection{The Radio-Aligned Component}

There have been a few hypotheses for the physical causes of extended UV
continuum emission in high redshift quasars and radio galaxies.  The most
viable ones are: 1) star formation stimulated by the radio jet as it
propagates outwards from the nucleus (McCarthy et al.~1987; Chambers et
al.~1987; De Young 1989; Rees 1989; Begelman \& Cioffi 1989), 2)
scattering of light from a hidden quasar by electrons or dust ({\it
e.g.,\ } Fabian 1989; Cimatti et al.~1997), 3) inverse Compton
scattering of microwave background photons by relativistic electrons in
the radio jets or lobes (Daly 1992a; b), and 4) selection effects
related to the possible enhancement of radio emission by dense gas
which is preferentially located along the galaxy's major axis (Eales
1992).  Observing that in fact the radio and UV continuum on small
scales are anti-correlated (see also Lehnert 1996), meaning that the
high surface brightness radio emission from the jet is actually in a
minimum in the rest-frame UV, provides a test of these various proposed
schemes.

In the model of jet induced star-formation and scattering hypothesis,
we might expect to see such anti-correlations on small scales in
addition to the ``alignment'' between the radio and UV continuum
emission.  This might come about for the same reason in both cases.
The pressure from the jet would push material outwards both along the
jet and perpendicular to it.  The high perpendicular pressure might
cause clouds to become unstable and collapse and it would also clear
material from the region of the jet.  In case of the jet-induced
star-formation hypothesis, overpressure due to the passing jet might
cause these clouds to form stars and in the scattering hypothesis, the
over-pressurized clouds would provide for more efficient scattering of
the quasar light.  In both hypotheses, this would explain the large
scale relationship ({\it i.e.,\ } the ``alignment effect'') but on
small scales ({\it i.e.,\ } the width of the jet) a anti-coincidence
whether generated by star-formation or scattering.  However, since it
is difficult to understand how the jet can inhibit stars from moving
into the regions through which it passes, the jet-induced
star-formation scenario only works if the crossing time of the high
mass stars is much longer than their evolutionary time scale.
Otherwise, the massive stars that are formed at the edge of the jet
will fill in the jet region with high surface brightness UV emission.
If the time scale for the massive stars to penetrate into the region of
the jet is long enough, the massive stars will die out, thus preserving
the ``hole'' in the light through which the jet is passing.
Interestingly, the radio source in PKS 2338+042 is likely to be young;
its observed small radio size (roughly a 10-20 kpc, modulo projection
effects) suggests that it is only between 10$^6$ and 10$^7$ years old
which is roughly the same order of magnitude as the evolutionary time
scale of high mass stars.  For example, if the stars are orbiting at
100 km s$^{-1}$, they will transverse 1 kpc (0.1\arcsec \ at the
redshift of the quasar in the adopted cosmology) in about 10$^7$
years.  The scattering hypothesis does not suffer from such a draw back
and is feasible with the only caveat that the jet must be fairly
efficient at removing possible scatterers from the regions through
which it is passing.  Given that the pressure in the jet is estimated
to be many orders of magnitude higher than the reasonable pressure of
the ISM in a normal galaxy (like the Milky Way for example), such a
possibility seems highly plausible.

The suggestion of Inverse Compton scattering seems to be ruled out by
these observations.  Inverse Compton scattering of microwave background
photons by relativistic electrons in the radio jets or lobes (Daly
1992a; b), would in fact require that the regions of the highest
electron density (likely to be the jets) should have the highest UV
surface brightnesses.  This is exactly the opposite of what we
observed.

Assigning the ``alignment effect'' to possible selection effects related
to the enhancement of radio emission by dense gas which is
preferentially located along the galaxy's major axis (Eales 1992) is an
interesting suggestion that seems plausible given the current data
set.  We have found evidence for ``jet cloud'' interactions in 2 out of
the 4 sources imaged at Ly$\alpha$ (see \S 5.3.1).  Therefore strong
interactions between the radio and ambient interstellar medium are
certainly not rare in radio galaxies or quasar hosts (\S 5.3 and
references therein).  Within this context, it may be that the ``hole''
in the UV continuum may be related to the increased pressure in the
region surrounding the jet due to the passage of the jet that in fact
prevents it from de-collimating.  However, in order to gauge whether or
not this speculation is plausible will require more extensive observations.

To make this discussion more generic, we note that other quasar hosts
appear to have this general alignment, but exhibit detailed spatial
anti-coincidence between the radio and rest-frame UV emission (although
perhaps not as dramatic as that seen in PKS 2338+042).  In HST snapshot
data on a large sample of quasars selected from the 3CR sample, Lehnert
(1996) found evidence for subtle anti-correlation between radio and
rest-frame UV continuum and line emission in these sources (also see de
Vries et al.~1996) even though generally the radio emission ``aligned''
with the rest-frame UV continuum and line emission.  These
anti-coincidences were mainly seen in the sources with complex compact
morphologies, roughly similar to the radio morphology of PKS 1318+113
and PKS 2338+042.

\subsubsection{A Possible Radio-UV Synchrotron Jet in PKS 0445+097}

As has been emphasized previously, an important issue in the study of
high redshift radio-loud AGN is the relationship between the
relativistic radio-emitting plasma and the line and continuum emission
from the host galaxies.  To test this hypothesis, we have overlaid a
0.16\arcsec \ resolution image of PKS 0445+097 from Lonsdale et al.~(1993) on
the F555W image (Figure 2).  Here we have assumed that the bright,
compact eastern component is identified with the quasar nucleus on the
basis of its inverted radio spectrum
($\alpha^{15GHz}_{5GHz}$=$-$0.6$\pm$0.2; Barthel 1984).  The overlay
shows that there is a curved optical feature southwest of the quasar
which corresponds to the radio jet.  The optical and radio flux densities
in a 0.5 $\times$ 0.5 arc second area centered on this feature are
0.18$\pm$0.04 $\mu$Jy ($\lambda_{obs}$=5398\AA, the center of the F555W
filter) and 13.5$\pm$0.7 mJy ($\lambda_{obs}$=2.0 cm), implying a
spectral index (S$_\nu$ $\sim$ $\nu^{-\alpha}$) of $\alpha^O_R$ =
1.1$\pm$0.2.  This is consistent with a steepening radio-optical
synchrotron spectrum since the radio spectral index of the jet between
5 GHz and 15 GHz is $\alpha^{15GHz}_{5GHz}$=0.8$\pm$0.2 (Barthel 1984)
and suggests that the emission may indeed be associated.  If this is
true, and deep HST imaging polarimetry would be required to prove this,
then the optical jet in PKS 0445+097 would be the most luminous and
highest redshift jet known (in the rest-frame of the quasar: log L$_B$
= 29.0 ergs s$^{-1}$ Hz$^{-1}$ and log P$_{1.4 GHz}$ = 34.6 ergs
s$^{-1}$ Hz$^{-1}$, more than an order-of magnitude more luminous than
any previously discovered synchrotron jet; see Dey and van Breugel 1994
for a discussion of known optical/radio synchrotron jets).  However,
considering the complex and filamentary structure of the circum-nuclear
host, there is of course also the possibility that the optical/radio
association is accidental, and the $\alpha^O_R$ fortuitously close to
the value expected for synchrotron emission.  We note also that the 5
GHz map of PKS 0445+097 published by Barthel (1984) shows a small
extension northeast from the core, {\it i.e.,\ } opposite to the
southwest radio jet, and coincident with the optical extension in that
same direction seen in the HST image.  No radio spectral index
information for this feature is available.

\subsubsection{Star Forming Regions in the Quasars Hosts?}

If the circum-nuclear UV continuum from the hosts are due to recent
star formation, than it is of interest to compare the UV luminosities
with low redshift starburst and normal galaxies (Kinney et al.~1993;
Donas et al.~1987; Treyer et al. 1998).  We find that the typical
luminosity of the circum-nuclear host is about 10$^{12}$ $L_{\sun}$
\ at $\approx$1700\AA \ ($\lambda P_\lambda$; Table 3), compared to
$\lesssim$ few $\times$ 10$^{11}$ $L_{\sun}$ for normal and starburst
galaxies at low redshift (H$_0$=50 km s$^{-1}$ Mpc$^{-1}$).  Thus the
host galaxies of quasars are at least an order of magnitude more
luminous than the most luminous low redshift galaxies in the UV.
However, Calzetti, Kinney, \& Storchi-Bergmann (1994) and Meurer et al.
(1997) find the typical UV extinction in the Kinney et al.~ sample and
starbursts generally to be about 1-3 magnitudes at $\approx$1700\AA.
If we correct the most extreme UV-emitting galaxies in the local
universe for this amount of UV extinction, they can indeed reach the UV
luminosities observed in the circum-nuclear hosts of the quasars.
Therefore analogs for the circum-nuclear host associated with these
quasars may exist in the local universe, but must be the most extreme
UV-luminous galaxies and, moreover, the quasar hosts must be relatively
unobscured.  There is no good reason to assume why the latter should be
the case.

To add some further perspective on the nature of the hosts of radio
loud quasars, we note that the UV luminosity of the hosts measured in
this study are much more luminous than the ``Lyman drop-out'' field
galaxies studied by by Steidel and collaborators (Steidel et al.~1996;
Giavalisco, Steidel, \& Macchetto 1996).  Using the UV (1500\AA)
luminosity function for Lyman drop-out galaxies presented by Dickinson
(1998), we find that the average UV luminosity of these 5 quasar hosts
is about a factor of 10 more luminous than the most luminous Lyman
drop-out galaxies.  This comparison was made without correcting either
sample for the effects of dust extinction.  Therefore, it is difficult
to associate quasar hosts with the field population of starburst
galaxies at high redshift -- even the extremely luminous ones.

\subsection{The Extended Ly$\alpha$ Emission}

Our quasars typically have Ly$\alpha$ luminosities of $\approx$few
$\times$ 10$^{44}$ ergs s$^{-1}$.  Ly$\alpha$ luminosities this high
are typical of what is observed in high redshift radio galaxies ({\it
e.g.,\ } van Oijk et al.~1997; McCarthy 1993).  This suggests that
quasars and high redshift radio galaxies have similar galaxy-scale
environments and ionizing sources, and supports models which attempt to
unify radio galaxies and quasars through orientation, evolution, and
environment.  In addition, even in our relatively short HST exposures,
the Ly$\alpha$ emission is extended over tens of kpc; again very similar
to what has been observed in high redshift radio galaxies.  These
results are in agreement with the conclusions of our other investigations
of high redshift (z$>$2) quasar hosts ({\it e.g.,\ } Heckman et
al.~1991a, b; Lehnert \& Becker 1998).

However, quasars offer us an advantage over the radio galaxies.  Since
we can observe the nucleus more directly, we can estimate the observed
ionizing flux emitted by the quasar nucleus.  Using the measured UV
fluxes from the HST data and using the scaling relations between the
flux of UV wavelengths and the total ionizing energy in quasars from
Elvis et al.~(1994), we estimate that in order to produce the
Ly$\alpha$ emission observed, the host galaxy intercepts only a few
percent of the total ionizing luminosity of the quasar.  This is not
surprising considering the observations of the ``proximity effect'' in
the Ly$\alpha$ forest lines.  The ``proximity effect'', where the
number density of Ly$\alpha$ forest lines is lower near the quasar than
away from it ({\it e.g.,\ } Weymann, Carswell, \& Smith 1981; Bechtold
1994), has been interpreted as the hydrogen becoming more highly
ionized within the sphere of influence of the quasar.  Observing this
effect implies that much of the ionizing radiation from quasars must
escape to cluster scales and thus the host galaxy of the quasar cannot
be completely optically thick to ionizing photons in all directions.
This is in agreement with our estimate that the host galaxy and
immediate environment of the quasar nucleus only intercepts a few
percent of the total ionizing luminosity.  Of course it could be less
since there could be local sources of ionization such as young stars
and/or shocks generated by the mechanical power from the radio jets. To
determine the relative importance of local ionization sources would
require spatially resolved spectroscopy.

The radio/Ly$\alpha$ overlays (Fig. 2) show that there is a good
association between the radio emission and the structure of the
Ly$\alpha$ emission.  First, the principal axis of the radio emission
is generally along the same direction as the extended Ly$\alpha$
emission.  This is very similar to the radio-aligned extended
emission-line regions seen in high redshift radio galaxies ({\it
e.g.,\ } McCarthy 1993 and references therein). Second, the surface
brightness of the line and radio emission appear anti-correlated to
some degree (see {\it e.g.,\ } Fig. 2).  This is generally seen at the
point where the jet and radio emission are curved. Third, it seems that
the brightest Ly$\alpha$ and radio emission are on the same side where
the radio lobe is closest to the quasars. This resembles the
radio/emission-line asymmetry correlation found for radio galaxies by
McCarthy, van Breugel, \& Kapahi (1991).  All of these properties may
be best understood as being due to strong interaction of the radio
sources (jets and lobes) with dense, asymmetrically distributed ambient
gas. Numerous examples of this are known in nearby radio galaxies ({\it
e.g.,\ } 3C277.3: van Breugel et al.~1985; 3C 405 and 3C 265: Tadhunter
1991; 3C 356: Eales \& Rawlings 1990; PKS 2152-699: Fosbury et
al.~1998; PKS 1932-464: Villar-Martin et al.~1998; 3C 171: Clark et
al.~1998).

\subsubsection{``Jet-Cloud Collisions''}

Our observations show two good examples in which jet-cloud collisions
seem to occur: PKS 1318+113 and PKS 2338+042.  In the case of PKS
1318+113 we observe two emission-line regions between which the radio
jet is passing. At this same location the radio jet bends. This
radio/optical morphology is very similar to that seen in some nearby
radio galaxies, especially Minkowski's Object (van Breugel et al.~
1985) and is strongly suggestive of a jet cloud interaction. In PKS
2338+042 the spatial resolution is insufficient to allow a similarly
detailed examination but the co-spatial bright Ly$\alpha$ and radio
knots east of the quasar and the radio jet curvature downstream from
this location suggest a similar jet-cloud collision occurs in this
object.

By analogy to the radio galaxies we will briefly examine whether the
observed cloud properties are consistent with such an interpretation.
With the limited data in hand (i.e., without high spatial resolution
spectroscopy) we can explore only few of the consequences expected from
a violent jet-cloud collision. The main questions we can address are 1)
are the clouds massive enough to deflect the jets, and 2) can they
survive the collisions on time-scales comparable to the radio source
ages?

The two Ly$\alpha$ blobs SW of the quasar nucleus in PKS 1318+113 have
L$_{Ly\alpha}$=1.1$\times$10$^{43}$ ergs s$^{-1}$ and
L$_{Ly\alpha}$=1.3$\times$10$^{43}$ ergs s$^{-1}$ to the east and west
of the radio emission respectively.  Assuming no dust, pure case B
recombination for 10,000 K (Osterbrock 1989), cylindrical or spherical
symmetry, and using the projected isophotal dimensions apparent in the
Ly$\alpha$ image as discussed previously, we then find that
$n_e^2f_V$$\sim$0.5 for both blobs, where $n_e$ is the electron density
and $f_V$ is the volume filling factor.  Since we do not have data to
estimate either $n_e$ or $f_V$ independently, we must rely on estimates
obtained for other objects. Rough estimates of the volume filling
factors for the extended emission line regions in various radio
galaxies suggest f$_V$ $\sim$ 10$^{-4}$ to 10$^{-6}$ and $n_e$ $\sim$
10 - 1000 cm$^{-3}$ (see {\it e.g.,\ } Baum et al.~1992; McCarthy 1993;
Lacy \& Rawlings 1994).  If for convenience we assume $n_e$ = 100
cm$^{-3}$, then this would imply a volume filling factor of about 5
$\times$ 10$^{-5}$ and thus consistent with values found by previous
studies.  These estimates would then imply a mass of ionized material
in these clouds of about few $\times$ 10$^7$ (n$_e$/100 cm$^{-3}$)
(f$_V$/10$^{-4}$) M$_{\sun}$.  Making similar assumptions for the
Ly$\alpha$ emission-line regions in PKS 2338+042, we find that
$n_e^2f_V$$\sim$1 and would thus estimate that the mass of the clouds
must be $\approx$ 10$^8$ (n$_e$/100 cm$^{-3}$) (f$_V$/10$^{-4}$)
M$_{\sun}$.  We note that the above estimates would be similar if we
assumed that the gas were shock heated instead of implicitly assuming
that the gas is in recombination equilibrium since most of the Hydrogen
line emission comes predominately from the post-shock recombination
zone ({\it e.g.,\ } Dopita \& Sutherland 1992).

Are such masses capable of deflecting the radio jets emanating for the
nuclei?  Theoretical modeling suggests that jets can be deflected by
discrete objects, but only if certain minimal criteria are met.  First
and foremost, the deflector must be sufficiently massive as so not to
be pushed out of the way of the radio jet too quickly.  Following the
arguments in Icke (1991) and McNamara et al.~(1996), we estimate that
deflecting clouds must have a mass,  $M_{cloud} \gtrsim 6\times10^4
M_{\sun} ({{0.1 }\over{\eta}})({{L_{rad}} \over{10^{42} ergs
\ s^{-1}}})({{c}\over{v_{jet}}})({{1 \ kpc}\over{l_{jet}}})({{t_{jet}}
\over {10^6 \ yrs}})^2$, such that it has not moved away by more than
the length of the undeflected jet, $l_{jet}$, during the lifetime of the radio
source, $t_{jet}$, and where $\eta$ is the conversion efficiency of the
jet power, $P_{jet}$, into radio emission such that $L_{rad}=\eta
P_{jet}$, and $v_{jet}$ is the velocity of the jet.  From the
characteristics of the radio emission given in Barthel (1984),
we estimate that the radio luminosity of PKS 1318+113 is about
4$\times$10$^{43}$ ergs s$^{-1}$ and of PKS 2338+042,
7$\times$10$^{43}$ ergs s$^{-1}$ and the ages of both sources is likely
to be of-order 10$^6$ to 10$^7$ yrs ({{\it e.g.,}}  Alexander \& Leahy
1987).  Using these estimates we see that the mass necessary to resist
being pushed out of the way is about 10$^8$ $M_{\sun}$ (assuming $\eta$
= 0.1, $v_{jet}$ = 0.2c, $l_{jet}$ = 1 kpc, and $t_{jet}$ = 10$^6$
\ yrs).  Our rough mass estimates above suggest that the clouds in PKS
1318+113 and PKS 2338+042 are sufficiently high for these clouds to be
able to resist the pressure of the radio jets for the relatively short
time that they have likely endured the passage of the radio jet.

Can the cloud survive the impact from a powerful jet?  In both sources
we observe what appear to be discrete clouds at the point where the
radio jets are deflected.  This suggests that a significant portion of
the clouds must have survived the impact in order to remain visible;
although perhaps the cloud in PKS 1318+113 has been split or had a hole
drilled through it.  The clouds, in absence of a confining external
pressure, will likely be disrupted on a time-scale of the order a few
times the sound crossing time.  Taking the cloud to have a gas
temperature of about 10$^4$ K, we estimate the sound speed $v_s$ = 15 km
s$^{-1}$. For cloud sizes of 3-4 kpc (which is the approximate
projected sizes of the clouds in our adopted cosmology), we then find
that the sound crossing time of these clouds is of-order 10$^8$ yrs.
This is much longer than the likely age of the jet, which is probably
comparable or smaller than the radio source age which, for powerful
double sources is estimated to be of order 10$^6$ to 10$^7$ yrs ({{\it
e.g.,}}  Alexander \& Leahy 1987).  Therefore we find, within the frame
work of our assumptions, that there is no particular problem with
seeing relatively intact, massive emission-line clouds millions of
years after the jet-cloud collisions occurred.

The above calculations suggest that 1) quasar hosts may contain rather
large amounts of dense clouds and 2) that these can be very efficient
at deflecting radio jets during a a significant fraction of the total
age of the source ({\it i.e.,\ } roughly 10$^7$ yrs).  To make this
arguments more general, we note that we only observed good evidence for
jet-cloud interaction in two sources, PKS 1318+113 and PKS 2338+042.
In two other sources, PKS 1658+575 and MRC 0549-213 we did not see
evidence for a jet cloud interaction, and in PKS 0445+097 we suspect
that there is something wrong with the narrow-band observation (\S
4.1.1.). Thus, we see strong cloud-jet interaction in two of the four
sources.  However, PKS 1658+57 and MRC 0549-213 (and also PKS 0445+097)
exhibit linear projected radio morphologies suggesting that no dense
clouds intercept the jets in these objects; although there is a region
of Ly$\alpha$ emission beyond the edge of the eastern radio lobe of MRC
0549-213 perhaps suggesting a large amount of confining material along
that direction.  Obviously a much larger sample of quasars with high
resolution Ly$\alpha$ images are necessary before as statistically
significant conclusion can be made.  However, our observations strongly
suggest that the bent radio structures in radio quasars may very well
be due to the interaction of their jets with dense ambient gas (Barthel
\& Miley 1988) and that such interactions may be very common and may
affect a large percentage of the total radio-loud high redshift quasar
population.

\subsubsection{Relevance to Unification Schemes}

The Ly$\alpha$ and continuum images of PKS 2338+042 and the Ly$\alpha$
image of PKS 1318+113 show relatively obvious signs of interaction
between the radio emitting plasma and ambient emission line gas and
perhaps even with the stellar population (as probed by the UV continuum
emission).  These results suggest that the interaction between the
radio and the ambient interstellar medium of the host galaxy and
cluster-scale environment must be important.  In fact, if we take our
data literally, they imply that interaction with the ambient medium is
important in determining the radio morphology in $\approx$1/2 of the
quasars.  Clearly, a much larger sample of high redshift quasars need
to be observed to determine the exact statistics of the impact of the
structure of the host galaxy and cluster-scale environment in
influencing the radio emission.

However, even for a limited number of quasars, this observation is
important.  Given that much of the evidence for orientation based
unification relies on various aspects of the radio morphology of the
sources (linear sizes, lobe arm length asymmetries, predominance of
jets, etc) this result allows us to speculate that one would not expect
there to be much evidence for orientation-based unification based on
radio observations alone.  {\bf IF} interactions between the radio and
inhomogeneities in the ISM of radio loud objects are important in
determining (and limiting) the size and lobe asymmetries in radio
sources, then such interaction might dominate over the simple growth of
the radio emission with time and orientation effects.  Studies that use
observations of linear sizes, lobe arm length asymmetries and bends and
twists in the radio jets and lobes to test unification schemes (e.g.,
Gopal-Krishna, Kulkarni, \& Whitta 1996; Kapahi et al. 1995) may in fact get
statistically insignificant results not because orientation-based
unification is incorrect, but because interacts between the
relativistic radio plasma and the ambient ISM and IGM either dominates
or provides a significant source of ``noise'' in the observations.
This may partially explain why the results of tests of unification
schemes using radio data have been so mixed and that orientation-based
unification seems most appropriate for a rather limited range of
redshifts and generally only for samples of relatively low redshift
radio sources (see Barthel 1989 for example).

\section{Summary and Conclusions}

In this paper, we presented HST WFPC-2 images of spatially-resolved
structures (\lq hosts\rq) around five high-redshift radio-loud quasars.
The quasars were imaged using the planetary camera with the broad-band
F555W (`V') filter and in the wide field camera with a narrow-band
filter whose central wavelength is approximately that of redshifted
Ly$\alpha$ in each of the quasars.  These radio-loud quasars were
selected from the earlier imaging survey of quasar \lq\lq host\rq\rq
\ by Heckman et al.~(1991a) and Lehnert et al.~(1992).  These HST images
confirm and extend our earlier ground-based results.

From an analysis of the images and a comparison with high resolution
VLA radio images we conclude that:

\noindent
1) All of the high redshift quasars are extended in both the
rest-frame UV continuum and in Ly$\alpha$.  We find extended fractions
that range from about 5\% to 40\% of the total continuum within a radius of
about 1.5\arcsec.  In spite of the fact that these images have higher
spatial resolutions and relatively short integration times on a small
telescope, the morphological agreement between the ground-based images
and these HST images is quite good.  Moreover, there is reasonable
agreement with our estimates of the fraction of the quasar light
contributed by the host galaxies in both the HST and ground-based
data.  Such a result is surprising given the fact that the HST images
reveal a wealth of structure within an arc second of the nucleus
which is currently unattainable from the ground.

\noindent
2) The typical integrated magnitude of the host is V$\sim$22$\pm$0.5.
The typical UV luminosity is roughly 10$^{12}$ $L_{\sun}$ ($\lambda
P_\lambda$, uncorrected for internal extinction), which is about a
factor of 10 higher luminosity than that observed for the ``Lyman drop
out'' field galaxies studied by Steidel and collaborators and the most
UV luminous zero redshift starburst galaxies.  The Ly$\alpha$ images
are also spatially-resolved. The typical luminosity of the extended
Ly$\alpha$ is about few $\times$ 10$^{44}$ ergs s$^{-1}$.  These
luminosities require roughly a few percent of the total ionizing
radiation of the quasar.

\noindent
3) Quasar host generally show ``alignment'' between the radio,
Ly$\alpha$, and UV continuum emission.  There is clear evidence that
the gas ``knows'' about the radio source.  This manifests itself in the
``alignment'' between the radio, Ly$\alpha$, and UV continuum emission,
in detailed morphological correspondence in some of the sources which
suggests ``jet-cloud'' interactions, and in the fact that the brightest
radio emission and the side of the radio emission with the shortest
projected distance from the nucleus occurs on the same side of the
quasar nucleus as the brightest, most significant Ly$\alpha$ emission.
These observations of jet-cloud interaction influencing the radio
morphologies is a challenge to simple orientation based quasar/radio
galaxy unification schemes.  This is perhaps why the use of the radio
morphology has generally lead to conflicting results when used to judge
the appropriateness of orientation based quasar/radio galaxy
unification schemes.

\noindent
4) The high spatial resolution of the HST has revealed that objects
along the line of sight but near the quasars in projection have made a
significant contribution to the continuum light from these objects.  We
note in particular that 0445+097 and 1318+113 were two quasars with
strong mis-alignment between the principal emission axes in the radio
and the ground-based images at UV rest wavelengths (Heckman et
al.~1991a).  It is now clear that this mis-alignment was partially due
to the contaminating effects of nearby (in projection; less than 2 arc
seconds for the quasar nucleus) foreground galaxies.

\acknowledgements
The authors wish to thank Ray Lucas for his considerable help in making
sure that our program went smoothly.  Conversations about the
complexities of the HST/WFPC2 PSF with Chris Burrows and John Krist
were particularly helpful in making the most of the data.  We thank the
referee, Dr. Eric Smith, whose comments lead to a substantial
improvement in the style and presentation of this paper and Dr. Greg
Bothun for his conscientious handling of the manuscript in his role as
the scientific editor.  The work of MDL and WvB at IGPP/LLNL was
performed under the auspices of the US Department of Energy under
contract W-7405-ENG-48 and the work of MDL at the Sterrewacht Leiden
was supported by funds provided by the Dutch Organization for Research
(NWO).  This work was supported in part by grant number GO-5393 from
the Space Telescope Science Institute, which is operated by the
Association of Universities for Research in Astronomy, Inc., under NASA
contract NAS5-26555.  We also acknowledge support from a NATO research
grant.

\newpage

\centerline{FIGURE CAPTIONS}

\figcaption [] {For each row of this plot, we show the F555W image with
the PSF removed as described in the text on the right and the
Ly$\alpha$ image is on the left, except for PKS 0445+097 for which we only
display the image taken through the F555W filter.  Each image has been
rotated so that north is at the top and east is to the left and each
has been smoothed using a 4 $\times$ 4 median filter.  The faintest
contours are listed in Table 2 and each contour is a factor of two
increase in surface brightness.  Each image is approximately 8\arcsec \
$\times$ 8\arcsec \ in projected size and the solid bar indicates
the angular size of 10 kpc at the redshift of the quasar assuming
H$_0$=50 km s$^{-1}$ Mpc$^{-1}$, q$_0$=0.1, and $\Lambda = 0$.}

\figcaption [] {We show the PSF subtracted F555W continuum or the
narrow-band (grey-scale) with contours of the high-resolution VLA 2 cm
radio map from Lonsdale et al.~(1993) overlaid.  In each of the plots
the radio nucleus as determined in \S 4 has been centered to the
position of the optical nucleus of the quasar.  The Figures are
specifically, a) F555W image of PKS 0445+097, b) F555W image of MRC
0549-213, c) narrow-band Ly$\alpha$ image of MRC 0549-213, d) F555W
image of PKS 1318+113, e) narrow-band Ly$\alpha$ image of PKS 1318+113,
f) F555W image of 1658+575 (4C 57.29), g) narrow-band Ly$\alpha$ image
of 1658+575 (4C 57.29), h) F555W image of PKS 2338+042, and i)
narrow-band Ly$\alpha$ image of PKS 2338+042.}

\figcaption [] {At the top, we show the comparison of a one-dimensional
light profile of the PSF subtracted F555W (rest-frame UV continuum)
image of PKS 2338+042 compared to the one-dimensional radio intensity
profile of the radio map from Lonsdale et al.~(1993).  Below, we show a
similarly constructed comparison of the narrow-band Ly$\alpha$ light
profile compared to the radio map of Lonsdale et al.~(1993).  Both
1-dimensional profiles are along right ascension which was chosen
because it lies directly along the radio jet.  The radio and HST images
were aligned and scaled for this comparison.  The flux scale is
arbitrary and chosen for convenience in making the comparison of the
light distributions at each wavelength.}

\newpage

\include{tab1}

\include{tab2}

\include{tab3}

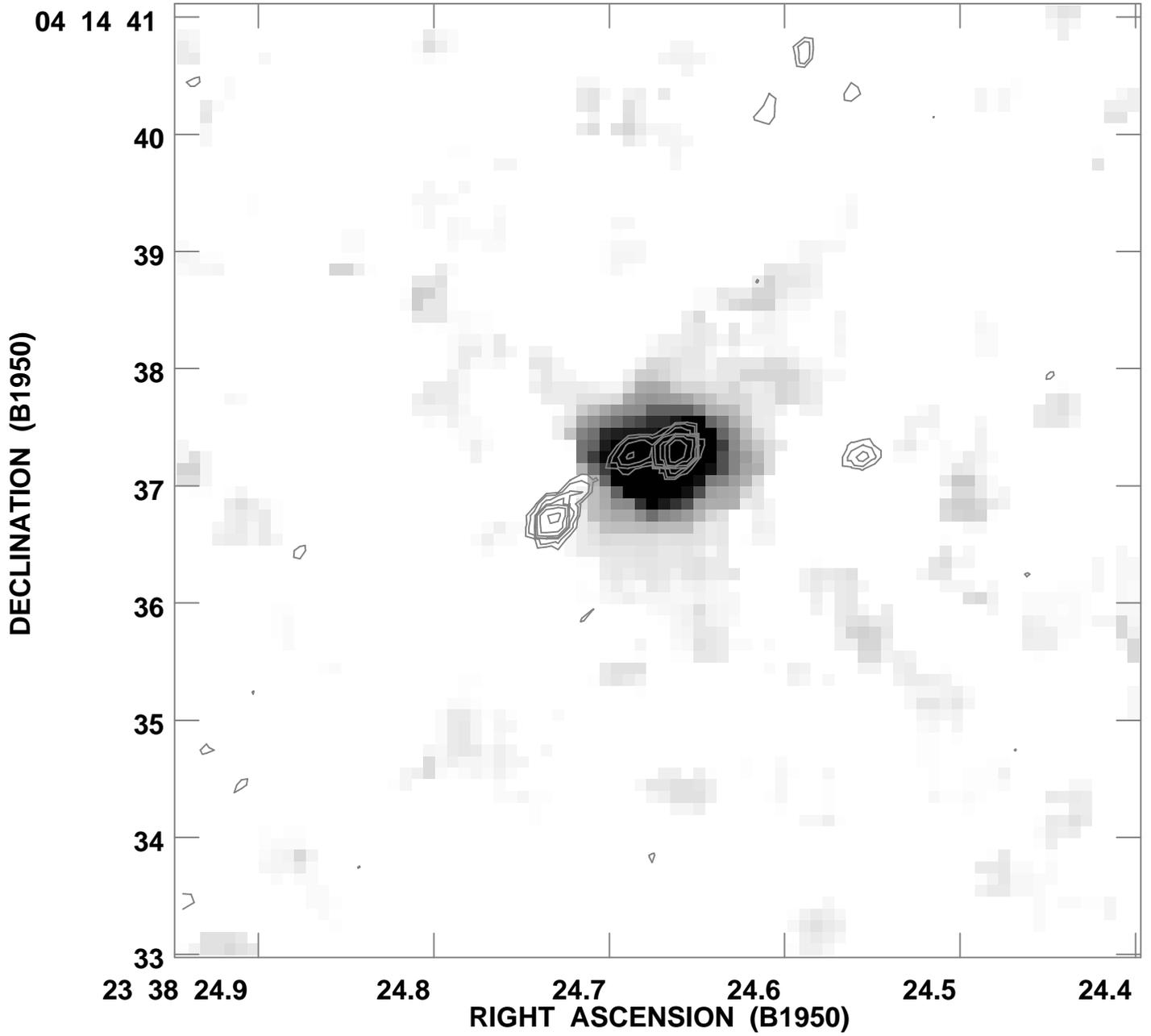
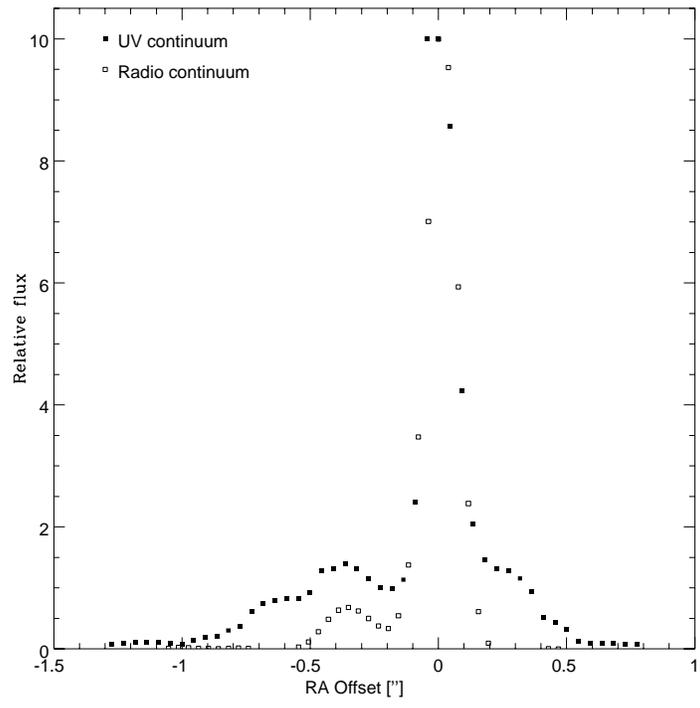
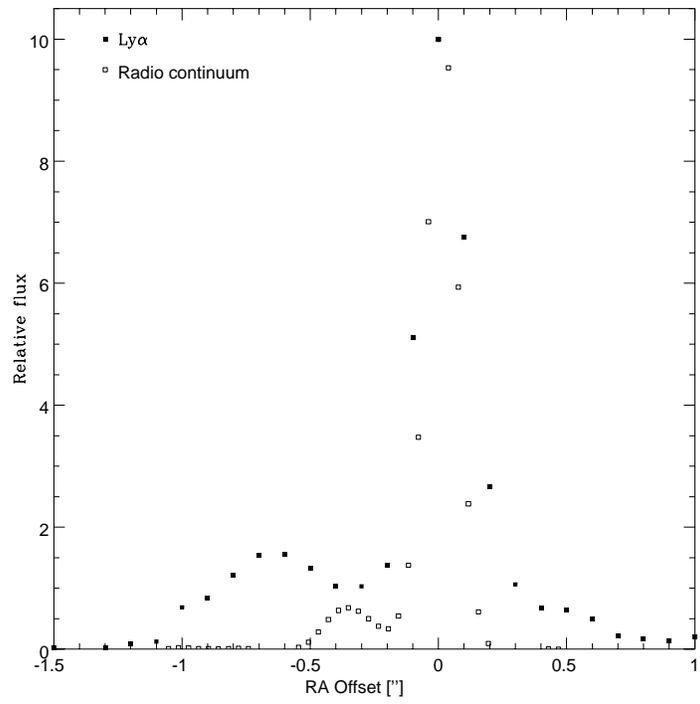
\end{document}

%% file: tab1.tex
\begin{deluxetable}{lccccc}
\tablecolumns{6}
\tablewidth{0pt}
\tablenum{1}
\tablecaption{Observation Log}
\tablehead{
\colhead{Quasar}&\colhead{Filter}&
\colhead{FOV}&\colhead{N}&
\colhead{Int}&\colhead{Date} \\
\colhead{(1)}&\colhead{(2)}&
\colhead{(3)}&\colhead{(4)}&
\colhead{(5)}&\colhead{(6)}}
\startdata
0445+097&F555W&PC&3&700&11/09/94 \nl
&FQUVN33&WP2&5&1000&11/09/94 \nl
0549--213&F555W&PC&3&700&22/04/95 \nl
&FQUVN-D&WF2&5&1000&22/04/95 \nl
1318+113&F555W&PC&3&700&23/05/95 \nl
&FQUVN-B&WF4&5&1000&23/05/95 \nl
1658+575&F555W&PC&3&700&12/01/95 \nl
&FQUVN-B&WF4&5&1000&12/01/95 \nl
2338+042&F555W&PC&3&700&06/12/94 \nl
&FQUVN-C&WF3&5&1000&12/01/95
\tablecomments{
Col. (1) --- Source designation.
Col. (2) --- Filter used for the observation.
Col. (3) --- The aperture used for the observation.  ``PC'' is
the planetary camera, and ``WF2'', ``WF3'', ``WF4'' are the wide
field cameras 2, 3, and 4.
Col. (4) --- Number of separate exposures each with the
exposure time in listed in Col. (5).
Col. (5) --- Integration time in seconds per exposure.
Col. (6) --- Date of observations in the form of day/month/year.
}
\enddata
\end{deluxetable}

%% file: tab2.tex
\begin{deluxetable}{lcccccc}
\tablecolumns{5}
\tablewidth{0pt}
\tablenum{2}
\tablecaption{Photometry of the Sources}
\tablehead{
\colhead{Quasar}&
\colhead{m$_{total}$}&\colhead{f(PSF-)}&
\colhead{Rotation}&\colhead{SB limit} \\
\colhead{(1)}&\colhead{(2)}&
\colhead{(3)}&\colhead{(4)}&
\colhead{(5)}}
\startdata
0445+097&20.2&0.19&$-$62.6&24.1 \cr
&$-$15.1&\nodata&26.9&\nodata \cr
0549$-$213&19.7&0.23&159.3&24.0 \cr
&$-$13.8&0.20&$-$111.2&-15.0 \cr
1318+113&19.0&0.38&173.7&24.0 \cr
&$-$13.8&0.18&84.1&-15.2 \cr
1658+575&18.3&0.21&9.5&23.0 \cr
&$-$13.5&0.16&$-$80.2&-15.2 \cr
2338+042&21.2&0.37&132.3&24.0 \cr
&$-$14.2&0.65&$-$47.8&-15.0 \cr
\tablecomments{
Col. (1) --- Source designation.
Col. (2) --- Total magnitude in the F555W filter or the logarithm of
the total Ly$\alpha$ flux (in units of ergs s$^{-1}$ cm$^{-2}$) of the
quasar.  The total magnitude of the quasar in the F555W filter is
listed in the first row for each object, while the logarithm  of the
total Ly$\alpha$ flux is enumerated in the second row.
Col. (3) --- Fraction of the total quasar light that is extended
as indicated by PSF subtraction.
Col. (4) --- Angle (in degrees) that the image was rotated to make
north at the top and east to the left in each image.  Positive values
imply a counter-clockwise direction of the rotation.  To calculate the
PA of the brightest diffraction spike one uses the following formula.
If the rotation is positive, its rotation $-$ 45$^\circ$, if the rotation
is negative, it is rotation $-$ 45$^\circ$ + 360$^\circ$. 
Col. (5) --- Surface brightness limit of the lowest contour of the
plots shown in Figure 1 in units of magnitudes arcsec$^{-2}$ for the
continuum images and in units of the logarithm of ergs s$^{-1}$
cm$^{-2}$ arcsec$^{-2}$ for the Ly$\alpha$ images.}
\enddata
\end{deluxetable}

%% file: tab3.tex
\begin{deluxetable}{lcccccccc}
\scriptsize
\tablecolumns{9}
\tablenum{3}
\tablecaption{UV Power and Ly$\alpha$ Luminosities}
\tablehead{
\colhead{Quasar}&\colhead{z}&
\colhead{$\lambda_{rest}$}&\colhead{f$_{\lambda, total}$}&
\colhead{f$_{\lambda, fuzz}$}&\colhead{log $\lambda P_{\lambda, tot}$}&
\colhead{log $\lambda P_{\lambda, fuzz}$}&\colhead{log L$_{Ly\alpha, total}$}&
\colhead{log L$_{Ly\alpha, fuzz}$} \\
\colhead{(1)}&\colhead{(2)}&
\colhead{(3)}&\colhead{(4)}&
\colhead{(5)}&\colhead{(6)}&
\colhead{(7)}&\colhead{(8)}&
\colhead{(9)}}
\startdata
0445+097&2.110&1740&$-$16.5&$-$17.2&12.6&11.9&\nodata&\nodata \cr
0549$-$213&2.245&1660&$-$16.3&$-$17.0&12.7&12.0&45.3&44.5 \cr
1318+113&2.171&1700&$-$16.0&$-$16.5&12.9&12.4&45.0&44.3 \cr
1658+575&2.173&1700&$-$15.8&$-$16.4&13.1&12.5&45.3&44.5 \cr
2338+042&2.549&1520&$-$16.5&$-$17.0&12.8&12.3&45.1&44.9 \cr
\tablecomments{
Col. (1) --- Source designation.  Col. (2) --- Redshift of the source.
Col. (3) --- The central wavelength of the F555W filter in the
rest-frame of the quasar using 5397\AA \ as the central wavelength of
the F555W filter.  Col. (4) --- Flux density of the total emission from
the quasars in units of the logarithm of ergs s$^{-1}$ cm$^{-2}$
\AA$^{-1}$ at the wavelength given in col. (3).  Col. (5) --- Flux
density of the ``fuzz'' in units of the logarithm of ergs s$^{-1}$
cm$^{-2}$ \AA$^{-1}$ at the wavelength given in col.  (3).  Col. (6)
--- The logarithm of the UV power of the quasar taken to be $\lambda
P_{\lambda}$ in units of solar luminosities using the rest wavelength
given in col (3) and the assumed cosmology of H$_0$=50 km s$^{-1}$
Mpc$^{-1}$ and q$_0$=0.1.  We have also corrected for galactic
extinction using the extinction in the B-band given in NED and using
the standard extinction curve from Osterbrock (1989).  The value of
solar luminosity used to make the conversion is $L_{sun}$ =
3.83$\times$10$^{33}$ ergs s$^{-1}$.  Col. (7) --- The UV power of the
``fuzz'' ($\lambda P_{\lambda}$) in units of the logarithm of solar
luminosities at the wavelength given in col. (3).  The relative
fraction of extended emission used in this calculation is from the PSF
subtraction analysis (see Table 2 and text for details).  Col. (8) ---
The total Ly$\alpha$ luminosity of the quasar in units of the logarithm
of ergs s$^{-1}$.  Col. (9) --- The total Ly$\alpha$ luminosity of the
``fuzz'' in units of the logarithm of ergs s$^{-1}$.  The relative
fraction of extended emission used in this calculation is from the PSF
subtraction analysis (see Table 2, cols. (2) and (3) for the fluxes
used and text for details).}

\enddata

\end{deluxetable}